%% file: JAlcolea.tex
\documentclass{evn2004}
\usepackage{txfonts}
\usepackage{graphicx}
\newcommand{\Msun}{\mbox{${M}_{\odot}$}}

\newcommand{\Lsun}{\mbox{${L}_{\odot}$}}

\newcommand{\my}{\mbox{${M}_{\odot}$~yr$^{-1}$}}

\newcommand{\s}{\mbox{$''$}}

\newcommand{\kms}{\mbox{km\,s$^{-1}$}}
\begin{document}
\include{page}
   \title{Circumstellar masers}

   \author{J. Alcolea}

   \institute{Observatorio Astron\'omico Nacional (OAN-IGN), C/ Alfonso XII 
              N$^{\underline{\rm o}}$ 3 y 5, E--28014 Madrid, Spain}

   \abstract{Circumstellar masers are unique tools for probing the
             envelopes around low mass dying stars using VLBI. In
             this contribution I briefly review the status of our
             knowledge on these emissions, especially focusing on
             the latest results obtained by means of VLBI observations.
             I also outline some of the new windows that will
             open for VLBI maser research thanks to the upcoming
             new instrumentation.
   }

   \maketitle
%

\section{Introduction}

Stars are tiny. In fact they are about the tiniest things one can find 
among astrophysical objects. The bigger a star can get is of the order
of several AU in the case of red super-giants, but (obviously) this size 
is just about 1\s\ at a distance of 1\,pc, and only 1\,m.a.s. at a more
typical distance of 1\,kpc. From this simple reasoning, it is easy
to see why the study of stars benefits from the use of the VLBI technique, 
which provides spatial resolutions of a fraction of a m.a.s.
at the highest available frequencies.

However, the observation of stars using VLBI is not so straight forward. 
VLBI provides very high angular resolution but we have to pay a price 
for it: we can only observe sources that at the same time are very compact 
and very bright. With the current instrumentation, VLBI flux 
sensitivities and resolving power imply lower detection limits for the 
brightness temperature from tens to hundreds of thousand degrees, in the case 
of continuum observations. The situation is of course much more restrictive
for spectral line studies, where minimum brightness temperatures
much higher  are required. For example, for a spectral resolution of a fraction
of \kms, the VLBA sensitivity at 7 and 3\,mm is about several 10$^8$\,K. In 
practice, these limits exclude the detection of {\em almost} any radiation 
of thermal origin, and typically VLBI studies deal with non-thermal sources: 
synchrotron and gyro-synchrotron emission in continuum studies, and maser 
emission and absorption in front of very strong non-thermal radio-continuum 
sources, in the case of spectral line works. However, I recall that behind 
these 
restrictions on minimum brightness temperature, there is no instrumental 
limitation other than sensitivity, and that relaxing those limits is just a 
matter of having more collecting area, better receivers, and, for the case 
of the continuum, broader simultaneous pass-bands. 

In any case, the current VLBI sensitivity restricts the number of stellar sources 
and/or processes that can be observed. In particular, we can hardly 
speak of detecting the stars themselves, only in the case of strong radio stars, 
but rather their ejecta and surrounding environments. In particular, VLBI studies 
of stars and related sources (proto-stars and star forming regions, and stellar ashes) 
include the observation of stellar magneto-spheres 
(M dwarfs, T Tauris, pulsars), of the high-speed ejecta of very compact sources
(black holes, neutron stars, white dwarfs), and the maser emission from 
several molecular species.

Maser emission occurs when there is an inversion in the populations of the two levels 
of one transition (i.e. the upper level is more populated than the lower level), leading
to an exponential amplification of the radiation that can attain extremely high brightness
temperatures.  These inversions are induced by asymmetries in the population/de-population 
processes of the energy levels connected by the maser transitions. 
(For example, in H$_2$O masers 
the upper levels of the transitions are mainly populated via collisions, 
while the de-excitation is due to radiative decays.) The maser emission itself tends to 
destroy the inversion, therefore, for the maser to maintain its power, a strong source 
of energy is required. This is the reason why masers are always associated to very 
luminous and energetic sources, AGNs, SNRs, high-mass proto-stars and Asymptotic Giant Branch 
stars. In the following sections I will give a general background on this last type of masers, 
also known as circumstellar masers, since they arise in the molecular envelopes around 
these stars. I will also highlight some of the latest results that have been obtained by 
VLBI observations of these strong emissions, focusing on their impact on aspects such as 
the maser pumping theory, the structure, dynamics and evolution of these 
envelopes, Astrometry and distance estimation, etc. Finally, I will outline how 
circumstellar 
maser studies may benefit from using the new and improved VLBI instrumentation, also 
in combination with other techniques, that is becoming or will become available within 
the next coming years. 

\section{Circumstellar envelopes} 

Stars with initial masses between 1 and 8 \Msun\ end their lives rather 
quietly, but also throwing away a large portion of the stuff they are made of. 
These mass losses occur mainly at the end of their evolution, while in the
Asymptotic Giant Branch (AGB) phase, 
and more specifically at the very end of it. Due to their moderate 
initial mass, after the Hydrogen and the Helium have been exhausted at the 
center of these stars, the core degenerates and the atomic reactions do not 
proceed in it any further. The energy of the star is mainly due to the CNO 
cycle occurring in layers outside the nucleus, which are surrounded by a deep 
convective mantle. The star becomes very luminous, \mbox{10$^4$ to 10$^5$\,\Lsun,} but 
at the same time very big, larger than 1\,AU, resulting in a red 
giant/super-giant, i.e. an AGB star. 

While in the AGB, stars are radial pulsators. These pulsations occur in
periods of one to two years, resulting in changes in the size (as large 
as 50\%) and temperature of the star. This is the reason why these objects are 
also known as Long Period Variables (LPVs). According to their brightness and
regularity of their variation, LPVs are divided in Miras (regular giants),
semi-regulars (semi-regular giants) and super-giants. 

During these pulsations, the atmosphere of the star rises and sinks at speeds 
of tens of \kms. These velocities are of the order of the escape velocity, 
and hence the material of the upper layers of the star is very weakly attached 
to it. The shock waves produced by the periodic pulsation form an extended 
atmosphere up to several stellar radii, where the density is still high but 
the temperature is below the condensation point for dust grains to form. Once 
dust grains are formed, they are kicked away by the impacts of the stellar 
photons. As the dust grains move out, they collide with the gas molecules and 
both gas and dust start a radial expansion thanks to the {\em radiation 
pressure} (provided that it can overtake the gravitational force). After a 
period of acceleration, and since both stellar radiation and gravity rapidly 
decrease as $1/r^2$, the material expands freely, forming a spherically 
symmetric envelope moving at a constant velocity of 10 to 30\,\kms. 
This process results in a mass loss that at the end of the AGB can attain rates 
as  high as 10$^{-4}$\,\my, and that will determine the future evolution of 
both the star and the envelope. See Olofsson (\cite{olofiau}) for a short review 
on AGB stars.

These circumstellar envelopes (CEs) are of course rich in dust and molecules, that 
can be observed mostly by means of the IR continuum emission (dust), and  
many spectral lines at cm, mm and sub-mm wavelengths (molecules).
The chemical composition of the envelope strongly depends on the composition
of the central star. According to their elementary abundances, AGB stars can
be divided in O-rich, with [O]/[C]\,$>$\,1, and C-rich, where [O]/[C]\,$<$\,1. There
is also a minor type in between, named S-type, for which [O]/[C]\,$\approx$\,1.
In addition, one of the most stable molecules in CEs is the carbon monoxide, CO.
For O-rich stars we can assume that CO forms until Carbon is exhausted in the gas phase.
This is the reason why O-rich envelopes are poor in C-bearing molecules other
than CO, and their dust grains mainly consist of silicates. On the contrary, C-rich
envelopes present lower abundances of \mbox{O-bearing} molecules, apart from CO, than
O-rich sources, and so their dust grains are formed of carbonaceous compounds.
See Habing (\cite{habingaar}) for a very complete review on both AGB stars and their 
CEs.

\section{Circumstellar masers}

Luckily, some of these circumstellar spectral lines are masers, allowing their 
study using VLBI. So far 5 molecular species have been found to be 
masing in circumstellar envelopes: OH, H$_2$O and SiO in envelopes of O-rich 
sources, and SiS and HCN in envelopes of C-rich sources. Very little is known 
on the masers of the C-rich stars, mostly based on single dish observations. 
They are detected just in a handful of sources, and to date no successful
VLBI observations of these masers have been reported. SiS shows maser emission 
in the $J$=1--0 line in the prototype source IRC\,+10216 (CW\,Leo). The line 
profile is U-shaped, the blue-shifted  component being more intense than the 
red-shifted one, see Nguyen-Q-Rieu et al. (\cite{rieu}). This could be a case 
similar to the OH masers, including some amplification of the stellar 
continuum on the blue peak (see Sect.\,3.1). HCN masers are detected in the $J$=1--0 
transition of the ground state (Izumiura et al. \cite{izumiura}) at 89\,GHz, and 
in three rotational lines of vibrationally excited states: the \mbox{$J$=1--0} line 
of the (02$^0$0) state (Lucas et al. \cite{lucas2}), the $J$=2--1 line of 
the (01$^{1\mathrm c}$0) state at 178\,GHz (Lucas \& Cernicharo \cite{lucas1}), and 
the $J$=9--8 line of the (04$^0$0) state at 805\,GHz (Schilke et al. 
\cite{schilke1}). More recently, Schilke \& Menten (\cite{schilke2}) have 
discovered maser/laser emission from the (ro-vibrational) cross-ladder 
(11$^1$0)--(04$^0$0) $J$=10--9 transition at 891\,GHz.  

Contrary to what happens in C-rich envelopes, masers in O-rich stars are very
widespread. They are detected in hundreds of sources and very complete statistical
studies exist. In addition, there are also many high spatial resolution
observations, either with VLBI or connected interferometry. Another important
aspect of the three species found to be masing in O-rich envelopes, OH, H$_2$O
and SiO, is that they are somehow quite complementary, since each of them arises
from the one of the three major parts we can distinguish in these envelopes:
the pulsating environment near the star (SiO masers), the regions were the
envelope is being accelerated by the impact of the photons onto the grains
(H$_2$O masers), and the outer envelope characterized by a constant expansion velocity
(OH masers). This certainly provides a very complete view of the envelope just
from the observations of these three types of masers.

\subsection{High precision Astrometry using OH masers} 

OH maser emission in CEs occurs in three different transitions at 18\,cm: 
the two main lines at 1665 and  1667 MHz, and the satellite lines 
at 1612 MHz, between levels in the ground $^2\pi_{3/2}$ $J$=3/2 state. 
In all cases, these OH masers are pumped via 
$^2\pi_{3/2}$--$^2\pi_{1/2}$\,$J$=3/2--5/2 transitions at 35\,$\mu$m, 
which absorb the radiation emitted by the circumstellar dust. 
OH masers are detected in the outskirts of the circumstellar envelope, at 
distances from the star of the order of 10$^{16}$\,cm. This location is due to 
the fact that OH is a very reactive radical and can only survive for some time 
in low density regions. In fact for OH to appear in the envelope, we need 
first to photo-dissociate the parent molecule, H$_2$O. Since the central star 
is cool, the photons required to break H$_2$O are provided by the 
interstellar UV radiation field. These photons efficiently penetrate in 
the outer layers of the envelope, where the dust density is low enough so not to 
block them. Of course the UV radiation also destroys the OH radical, and so the 
masers are confined to the layers between those of the photo-destruction of 
H$_2$O and OH itself. Main lines are dominant in stars with low mass loss rates.
As the envelope becomes thicker, 1612\,MHz masers take over, but for very thick
envelopes the 1667\,MHz line becomes dominant again. The line profiles of
the OH masers are very characteristic: they are U-shaped. This profile 
is simply because OH masers arise from a thin shell that is radially 
expanding at an almost constant velocity (of some tens of \kms, much 
larger than the turbulence velocity of the gas). Because of this particular 
velocity field and geometry, for any observer, the light trajectories for 
which the velocity coherence path (and maser amplification) is the largest, 
are those passing near the star, i.e. both most blue-shifted and red-shifted
ones. This model was confirmed with the first observations of OH masers using 
MERLIN (Booth et al. \cite{booth}) and the VLA (Baud \cite{baud}). 

Since the masers are pumped via the IR radiation of the star, the OH masers
also vary with the same period (Herman \& Habing \cite{hyh}). A delay is 
observed between the maxima for the blue and red peaks, simply due to 
the extra path that the light coming from the receding part of the envelope needs 
to travel. In fact, this delay can be used to measure the diameter of the emitting 
shell $d$, which is given by $d=ct$, where $c$ is the speed of light, and $t$ is 
the measured delay of the red-shifted peak w.r.t. the blue-shifted one (Jewell et 
al. \cite{jewell}). In addition, the diameter of the shell in the plane of the sky 
measured in angular units $\theta$, is just the diameter of the emission at the 
systemic velocity (at the mid point between the two peaks). Assuming that 
the shell is spherically symmetric, we obtain that the distance to the star $D$, is 
given by $\theta = d/D$ (see Spaans \& van Langevelde \cite{spaans} for further 
details).
  
OH maser emission is too big in stars for deserving VLBI observation just from 
cartographic consideration. In fact, most of the emission is resolved out when
observed with long baseline interferometry. However, the VLBI observation of 
this emission has a potential impact in Astrometry studies as Vlemmings et al. 
(\cite{vlemmings}) have recently shown. Following the model described before, 
it is clear that the most blue-shifted emission  should correspond to the line of 
sight passing through the star. It is then possible that this blue peak would be 
also amplifying the continuum of the star itself. If this is the case, the 
astrometric measurement of such a spot would directly give us the accurate position 
of the star behind. Repeating this measurement for a period of several years, we 
will obtain the proper motion of the star superimposed on the movement
due to the annual parallax. Provided that we can separate both effects on the 
measured trajectory of the star/maser spot on the sky, we will directly
obtain the distance to the target. This technique has been used
by Vlemmings et al. in two sources, U\,Her and W\,Hya, obtaining results in 
agreement with Hipparcos distances. In two more cases, R\,Cas and S\,CrB,
they used a compact red spot, hoping that its relative motion w.r.t. the
star is small. In these cases the obtained distances were almost
consistent with Hipparcos data.

\subsection{Measuring the acceleration of the envelope using water vapor masers}

Moving inwards to denser regions in the envelope, the next masing region we 
find is that of the 22\,GHz H$_2$O masers, corresponding to the 
6$_{1,6}$--5$_{2,3}$ rotational transition in the ground vibrational state. 
The 22\,GHz line is not the only transition of H$_2$O showing maser action, but 
it is certainly the best studied and observed among water masers, and the 
only one for which high spatial resolution images are available. For the other 
maser lines very little is known. In addition to the 22\,GHz line, there are 
circumstellar water masers detected in the \mbox{3$_{1,3}$--2$_{2,0}$}
(183\,GHz), 10$_{2,9}$--9$_{3,6}$ (231\,GHz), and 5$_{1,5}$--4$_{2,2}$ 
(325\,GHz) lines from the ground vibrational state (see Cernicharo et al. \cite{cerni}, 
Menten et al. \cite{menmelphi}, Menten \& Melnick \cite{menmel0}, and Yates 
et al. \cite{yatesetal}). Models for these maser lines predict that they are
inverted more or less in the same shells of the envelope than the 22\,GHz masers. 
There are also water vapor maser lines arising from 
the first vibrationally excited state of water $\nu_2$=1 (the bending mode): 
4$_{4,0}$--5$_{3,3}$ (96 GHz), 5$_{5,0}$--6$_{4,3}$ (233 GHz), and 
\mbox{1$_{1,0}$--1$_{0,1}$} (685 GHz), see Menten \& Melnick (\cite{menmel1}), and 
Menten \& Young (\cite{mentaco}). 
Since the excitation of these masers is much higher, it is 
assumed that they are originated in much more inner regions, like
in the case of SiO masers (see Sect.\,3.3).

The inversion mechanism of the ground state water masers is well understood. 
The upper level of the maser transitions always lie along (or near) 
the {\em backbone} (levels with $J_{1,J}$ or $J_{0,J}$ quantum numbers).
Radiative transitions between levels along the backbone are very strong, and
become optically thick at lower densities/column densities. Because of these
high opacities, the collisional excitation of the H$_2$O molecules results 
in higher populations for backbone levels, leading to inversions w.r.t. 
levels somewhat lower in energy but away from the backbone 
(see e.g. Neufeld \& Melnick \cite{neufeld}). For explaining the 
pumping/inversion of the  vibrationally excited masers there is no satisfactory 
model yet (Alcolea \& Menten \cite{alcomen}).

VLA observations of 22\,GHz water masers (see e.g. Bowers \& Johnston 
\cite{bowers}, and Colomer et al. \cite{coloagua}), show that they
arise from regions of the CEs with sizes of the order of 10$^{15}$\,cm, 
at distances to the central stars up to about 10 stellar radii. 
These works, although were unable to resolve the individual spots, indicate 
that in these regions of the envelope the gas is radially expanding but it 
has not yet reached the terminal velocity, i.e. the gas and dust are still 
being accelerated by the radiation pressure. 

These results have been greatly
improved by the works carried out with MERLIN which, thanks to its improved 
spatial resolution over the VLA, is able to resolve the individual maser
spots, allowing a much better study of the water masers and the circumstellar
shells from which they arise. Yates \& Cohen (\cite{yatescohen}) 
identified the first proper motions of water masers, and also
demonstrated that the size of the emitting region increases with the 
mass loss rate, as predicted by the models of Cooke \& Elitzur (\cite{cooke}). 
These proper motions have been measured now in a number of objects: NML\,Cyg 
(Richards et al. \cite{nmlcyg}), VY\,CMa (Richards et al.  
\cite{vycma}), S\,Per (Marvel \cite{marvel}), and VX\,Sgr (Murakawa et al. 
\cite{vxsgr}, Marvel \cite{marvel}), all these four stars being super-giants.
The results from the proper motion studies of the 22\,GHz line are very
consistent. The thin shell model which expands at constant velocity
does not apply here. On the contrary, water vapor masers arise from a
thick shell, some times not spherically symmetric. This shell is 
in radial expansion, but with a velocity that increases by a factor about two 
between the inner and outer limits of the maser region.
This implies values of $\epsilon$, the logarithmic velocity gradient, 
between 0.5 and 1.0.

Bains et al. (\cite{bains}) have recently published MERLIN observations
of water masers in other types of AGB sources: the Mira variables
U\,Her, U\,Ori and IK\,Tau, and the semi-regular variable RT\,Vir.
Although in this case no proper motions are available, the velocity
vs. position plots show that also in these cases, the radial velocity
field has a clear outward gradient. It seems then that all these 
observations are given us information on the kinematics of the envelope
when there is still some physics on it (other than free expansion).
In fact, today there is no other way in which we can directly probe
the velocity field in these regions of the envelope. In addition, when
the (multi-epoch) observations include proper motion measurements, 
estimations for the distance to the star can be done assuming a simple
structure for both the envelope and its velocity field. The results of 
these values for the distance are in fact in good agreement with those
derived using other methods.

Very briefly, I would like not to forget the works by Imai et al. 
(\cite{imai02} and \cite{imai04}). These authors have also computed
proper motions of the H$_2$O maser spots in circumstellar envelopes
but of two early post-AGB stars, showing that in these
cases water masers trace very young molecular bipolar outflows,
that characterize this type of sources (see Sect.\,4.).

\subsection{Multi-transitional studies of circumstellar SiO masers}
 
Circumstellar SiO masers are detected in the three main isotopic 
substitutions of SiO, $^{28}$SiO, $^{29}$SiO, and $^{30}$SiO, in rotational 
transitions from the $J$=1--0 (at 43 GHz) up to the $J$=8--7 at (350 GHz)
of the fundamental $v$=0 and $v$=1 to 4 vibrational excited states. 
SiO masers are located in the innermost layers of the circumstellar
envelopes. This was soon suggested simply because of their high excitation 
requirements (the $v$=1 states of SiO are 1770\,K above the $v$=0), from the 
shape of their spectra, and the fact that no large abundances of SiO are 
expected in the layers outside the dust formation layer. (The silicate 
dust grains in O-rich stars are in fact made of gaseous SiO after nucleation 
and condensation). However the exact location of the SiO maser remained 
unknown until the first VLBI true maps of the $v$=1 $J$=1--0 line showed that 
they consisted in a chain of spots delineating a ring of just a few stellar 
radii in diameter (Diamond et al. 1994). Although the position of the star in 
the map is not known (and this is still the case for all VLBI maps of SiO 
masers), the ring geometry strongly supports that the star should be located 
near the centroid of such structure.

The basic principle for explaining the inversion of SiO masers (for 
vibrationally excited transitions only) is very simple. It was 
proposed by Kwan \& Scoville (1974), and it is based on the 
{\em self-trapping} phenomenon. For a SiO molecule in a vibrationally 
excited estate, the main de-excitation route is the decay, via ro-vibrational 
transitions, to lower vibrationally excited levels. When these transitions are 
optically thick, the escape probability of these photons is inversely 
proportional to the opacity of the line, and hence to the $J$ quantum number of 
the upper state. Because of this, the higher the $J$ value the less probable
for a molecule to de-excite, resulting in a chain of inverted populations
along the rotational ladder. Of course, this mechanism only works provided that
the inverted levels are populated at similar rates, i.e. not depending on the
$J$ number. This general inversion mechanism can be strongly modified 
by overlaps between the ro-vibrational lines of the SiO species and of 
other abundant molecules like H$_2$O (see e.g. Olofsson et al. 
\cite{olofsson}, and Gonz\'alez-Alfonso \& Cernicharo \cite{eduardo}).

As for the source of energy responsible for the pumping of the masers, there is 
still a debate after more than 30 yr since the discovery of these masers, and 
almost 14 yr since their first successful VLBI observations (Colomer et al. 
1992). There are two main types of pumping mechanisms, termed {\em radiative} 
or {\em collisional} depending on the main source of energy responsible for 
the pumping. In collisional models, see e.g. Gray \& Humphreys (\cite{gray}) and 
Humphreys el al. (\cite{humphreys}), the SiO molecules are pumped to $v$$\ge$1 
states via 
collisions with H$_2$, whereas in radiative models, see e.g. Bujarrabal (\cite{buja1}
and \cite{buja2}), the energy is obtained directly from the stellar 8\,$\mu$m 
IR radiation (the wavelength of the SiO $v$=1--0, 2--1, etc. ro-vibrational 
transitions). 

\begin{figure*}
   \centering
      \includegraphics[width=\textwidth]{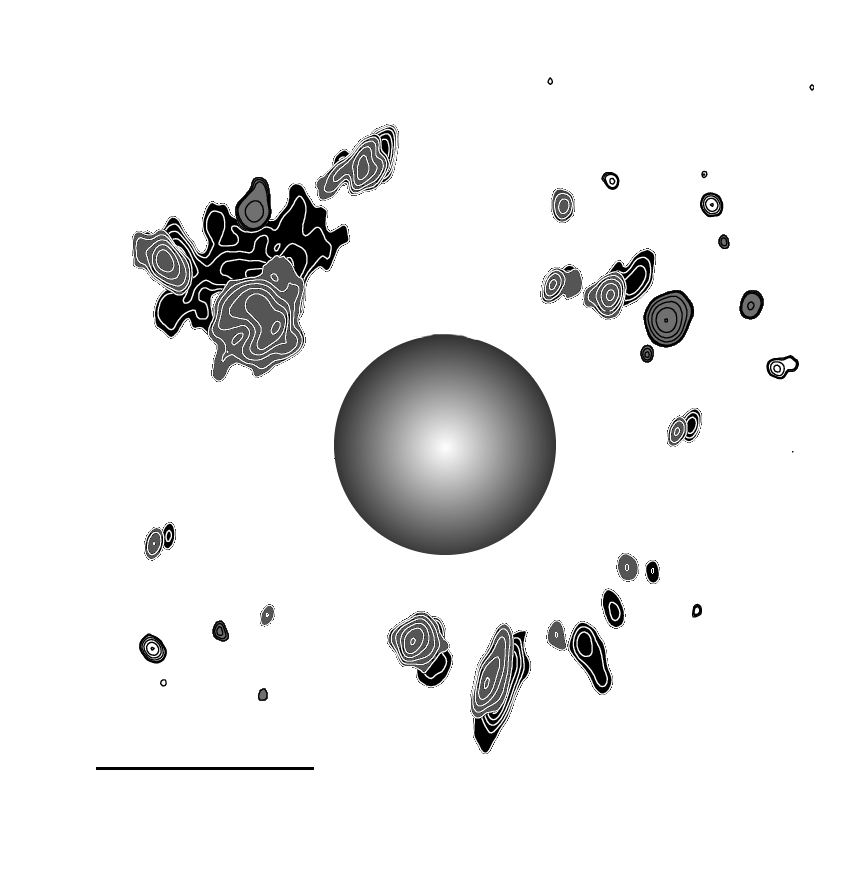}
      \caption{Distribution of SiO masers in the innermost shells of
        the circumstellar envelope around the long period variable
        IRC\,+10011 after Soria-Ruiz et al. (\cite{rebeca2}). These
        authors observed four different maser lines, $^{28}$SiO $v$=1
        $J$=1--0 (black with white contours), $^{28}$SiO $v$=2 $J$=1--0
        (grey with white contours), $^{28}$SiO $v$=1 $J$=2--1 (white
        with black contours), and $^{29}$SiO $v$=0 $J$=1--0 (grey with
        black contours), with excitation temperatures from zero to more
        than 3500\,K.  To relatively align the different maps, the
        centroids of all the transitions have been assumed to be at the
        center of the star. The diameter of the star, 11 m.a.s., is
        taken from the model by Vinkovic et al.  (\cite{vinkovic}).
        The horizontal black line has a length of 10 m.a.s. in the
        scale of the plot (which translates into a linear size of
        7.3\,10$^{13}$\,cm for a distance to the source of 500\,pc).
        The maser spots clearly delineate a ring, approximately located
        between 1 and 2 stellar radii.  In this object, it has been
        measured that the first layers of thick dust are located at a
        radius of 33 m.a.s., about 6 stellar radii (Lipman et
        al. \cite{lipman}).  }
        \label{fig-irc}
   \end{figure*}

Both models have pros and cons. On the one hand, for the radiative model to 
work, it is required that the absorption of the 8\,$\mu$m radiation is done 
under optically thin conditions, but I have just said that to attain the 
inversion these radiative de-excitations need to be optically thick. 
This situation is not contradictory if the SiO maser region presents a long 
velocity gradient, or has a thin shell geometry. Radiative models predict that 
SiO masers should show preferentially tangential amplification, i.e. the observed 
ring shapes, and that their intensity should vary in phase with the IR flux of the 
star, as it has also been observed (Alcolea et al. \cite{alcosegui}, Pardo 
et al. \cite{pardo}). On the other hand, collisional models do not need 
any especial geometry or velocity field for producing the population inversion,
but they do need them for explaining the observed ring like distributions. 
However, the main drawback of these collisional models is that it is very 
difficult to explain how the maximum flux of all maser spots (including different 
maser transitions) occurs within less than one month. For example, for a maser ring 
thickness of 4.5\,10$^{13}$\,cm (i.e. 3\,AU, or 6\,m.a.s. at a distance of 500\,pc, 
see Fig.\,\ref{fig-irc}), we need a physical process traveling at speeds larger 
than 90\,\kms\ to affect all the maser spots in less than 60 days. 

The observation of a single SiO maser line does not help very much in supporting
one model against the other, since both can predict the observed source size 
and maser strength. However, it is expected that the observations of several 
lines, with similar and different excitations, impose more severe constrains to
the models, helping in determining which is the correct (if any). These 
multi-transitional observations need to be (almost) simultaneous, since SiO 
masers vary both in intensity and location in the shell 
(Diamond \& Kemball \cite{diamond2}). During the last century, VLBI 
observations of SiO masers concentrated on a single line, the 
$v$=1 $J$=1--0 of $^{28}$SiO in their majority, but in the past few years 
simultaneous observations of two, three, and up to four $^{28}$SiO 
maser lines have become available.

The comparisons of the distributions of the $v$=1 and $v$=2 $J$=1--0
masers have reached, I think, a firmly established result. As it was
first pointed out by Desmurs et al. (\cite{desmurs}), both masers show
distributions which are very much alike. It is true that both
transitions share some emitting regions, but for the majority of the
spots there is a systematic shift of about 1\,m.a.s. between the two
maser lines, the spots of the $v$=1 being always farther away. This
result holds regardless of the optical phase of the star (Cotton et
al. \cite{cotton}). Such a small difference seems very surprising for
two lines so separated in excitation (1770 vs. 3540\,K) and this is
very difficult to explain using the standard models for SiO masers.
This problem becomes even more severe when considering also the
comparison between the $v$=1 $J$=2--1 and $J$=1--0 masers. So far we
have only data of such a comparison for three sources, but the results
are shocking. In the case of IRC\,+10011, Soria-Ruiz et al.
(\cite{rebeca1} and \cite{rebeca2}, see also Fig.\,\ref{fig-irc}) have
obtained for two epochs that, in spite of being two adjacent
transitions, the differences between the two $v$=1 lines are much
larger, 4\,m.a.s., than those for the two \mbox{$J$=1--0} lines (which,
as I mentioned before, are 1770\,K apart). From inside to outside they
find the rings of the $v$=2 $J$=1--0 line, then the $v$=1 $J$=1--0, and
finally the $v$=1 $J$=2--1.  The same series of rings have also be
found by Winter et al.  (\cite{winter}) in R\,Cas in three different
epochs.  (A fourth epoch of R\,Cas has been published by Phillips et
al.  \cite{phillips}, but since these authors did not mapped the $v$=2
$J$=1--0, we can not examine whether the $v$=1 $J$=1--0 is more like
the $v$=2 or like the $v$=1 $J$=2--1.)

As pointed out by Soria-Ruiz et al. (\cite{rebeca1}), these results are
incompatible with current pumping models for SiO masers, since all
predict very similar spatial distributions for lines within the same
vibrational state, and different ones for lines with very different
excitation. These authors propose that the line overlap between
ro-vibrational transitions of H$_2$O and SiO may solve the problem. In
fact they conclude that, by introducing the effects of the extra
opacity in the $v$=2--1\,$J$=1--0 ro-vibrational transition of
$^{28}$SiO due to the $\nu_2$=1--0\,11$_{6,6}$--12$_{7,5}$ of H$_2$O,
the distribution of the masers in IRC\,+10011 can be explained (see
also Bujarrabal et al. \cite{buja-over}).

This hypothesis is also supported by the results obtained for
$\chi$\,Cyg, where Soria-Ruiz et al. (\cite{rebeca1}) also detected the
$v$=2 $J$=2--1 maser. In this case the two $v$=1 masers delineate rings
of about the same size, whereas both $v$=2 masers are much more
compact, just as expected from classical pumping models. However
$\chi$\,Cyg is a S-type star, for which H$_2$O abundances lower than
those in O-rich sources (like IRC\,+10011 and R\,Cas) are expected. In
fact, neither H$_2$O nor OH masers have been detected in this source
(Benson et al. \cite{benson}), and therefore it seems very probable
that the effects of the proposed overlap could be much weaker here.

These are of course very preliminary results, that need much further 
investigation (both from the observational and theoretical point
of view), but my understanding is that from now on we should face the fact 
that for modeling SiO masers in circumstellar envelopes, 
the effects of the line overlaps can not be neglected.
I must also note this {\em anomalous} shift between the $v$=1 $J$=1--0 
and $J$=2--1 masers of $^{28}$SiO has also been reported by Doeleman et 
al. (\cite{doeleman}) in Ori\,A\,Irc\,2. 
 
\section{Rotation and magnetic fields in post-AGB envelopes and before}

As I said before, at the end of the AGB, the {\em super-wind} phase,
the mass loss rate can be as large as 10$^{-4}$\,\my. Obviously, the
star can not endure such an enormous mass loss rate for a long
time. After a few ten thousand years the H/He-burning shells become
literally exposed, and the star turns extremely hot and tiny becoming a
blue dwarf. This transformation occurs very rapidly, lasting no more
than a few thousand years. Meanwhile, the circumstellar envelope is
deeply changed. The heavy mass loss has stopped, and the CE becomes
thiner and thiner as it keeps expanding, being first photo-dissociated
and then photo-ionized by the hard UV radiation emanating from the hot
central star (once it becomes hotter than 30,000\,K), leading to the
formation of a Planetary Nebula (PN). The shaping of the PN is also due
to the interaction of the slow wind from the AGB phase (the former CE)
with a more tenuous, but faster and hotter one, that is released by the
central star at this stage. This Interacting Stellar Wind (ISW) model
was proposed by Kwok et al. (\cite{isw}) to explain the morphology of
spherical PNe. However, this interaction can only produce spherically
symmetric PN. Nowadays there is mounting evidence that in general PNe
are anything but spherical, and therefore modifications to the ISW
model are required.

Traditionally for explaining non-spherically symmetric PNe it has been
assumed that during the super-wind phase, at the end of the AGB, the
mass loss is not isotropic but somehow enhanced along certain
equatorial plane. The interaction of this non-isotropic AGB wind with
the isotropic wind in the PN phase would naturally result in the
formation of asymmetric PNe: this modification of the ISW model is
known as the GISW (Generalized ISW) model (Mellema \& Frank
\cite{gisw}).  However there is a major problem: in general, AGB
envelopes have little deviation from the spherical symmetry, as one
would expect for an isotropic loose of mass. This has been proved by
means of observations of the CO emission (Neri et al. \cite{neri}), OH
and SiO masers (see previous sections), and the detection of concentric
rings/arcs of dust scattered light around AGB and post-AGB sources (see
Balick \& Frank \cite{balick}).

Today we know that the shaping of these asymmetrical PNe is probably
due to the interaction of a fast collimated wind, released at the very
end of the AGB phase, which collides with the slow moving AGB
wind. These bipolar flows are observed by means of high velocity
emission in molecular and atomic lines of post-AGB stars, whose
envelopes have not yet attain the PN phase, the so called pre-PNe
(PPNe). About what powers these jets the only thing we know is that
radiation pressure is not responsible for them (Bujarrabal et
al. \cite{bujalas}, see also Alcolea \cite{zermat} for a review on the
characteristics of these bipolar ejecta, and their importance in the
shaping of PPNe and PNe).  Not having other candidates, and based also
in the universal connection between bipolar flows and rotating strong
magnetic fields (AGNs, micro- and nano-quasars, proto-stars, pulsars),
some authors have postulated that a process similar to the
magneto-centrifugal mechanism may operate in late AGB and post-AGB
sources.

The best way of having a look at any possible rotation in CEs using
VLBI is to observe SiO masers. Since they are the closest to the star,
it is expected that they show larger rotation velocities. So far, the
detection of rotation from VLBI observations of SiO masers has been
claimed only in four stars.  Hollis et al. (\cite{hollis}) have found
that the SiO masers in R\,Aqr are in almost Keplerian
rotation. However, this case is somewhat special since it is a well
known symbiotic system displaying a hourglass nebula.
S\'anchez-Contreras et al. (\cite{carmen}) also found that positions
and velocities of the SiO maser spots in OH\,231.8+4.1 could indicate
rotation.  Again OH\,231.8+4.2 is not a normal example since it is
highly bipolar post-AGB source (see Alcolea et al. \cite{oh231co}).
Boboltz \& Marvel (\cite{nmlrot}) also reported rotation from the SiO
masers in the super-giant NML\,Cyg. However they assumed a systemic
velocity of $-$6.6\,\kms, whereas from thermal CO emission the systemic
velocity is about $-1$\,\kms\ (Kemper et al. \cite{kemper}).  Cotton et
al. (\cite{cotton}) reported indications of rotation just in R\,Aqr and
possibly in S\,CrB, out of the nine targets observed.  In summary,
excluding {\em peculiar} cases, of about a dozen stars mapped in SiO
maser emission, there may be indications of rotation only in two of
them. Therefore, it seems that rapid rotation is not a common feature
to AGB stars, but further investigation is required.

Polarization measurements of the different masers can tell us about the
strength of the magnetic field in circumstellar envelopes. SiO masers
usually show linear polarization tangential to the circles they
delineate, i.e. perpendicular to the radial direction (see e.g. Kemball
\& Diamond \cite{kemball}, and Desmurs et al. \cite{desmurs}). However,
this is a natural result if the masers are pumped via the stellar
radiation (since the linear polarization vector must be orthogonal to
both the direction of the amplification of the masers and of the
pumping radiation). Circular polarization of SiO masers is less
frequently observed. Kemball \& Diamond (\cite{kemball}) reported its
detection in TX\,Cam, implying (according to these authors) magnetic
fields of 5--10\,G for the SiO maser region.  The situation is more
clear for OH and H$_2$O masers.  The works on OH masers by Szymczak \&
G\'erard (\cite{szymczak2004}) and Bains et al. (\cite{bains2003} and
\cite{bains2004}) are systematically finding polarization properties
consistent with magnetic fields of the order of a few mG. The studied
sources are always PPNe or candidates to be PPNe, and in some cases it
seems that the non spherical symmetry of the nebula could be related to
the structure of the magnetic field (see also Etoka \& Diamond
\cite{etoka}, and Szymczak et al. \cite{szymczak2001}). Very recently
Vlemmings et al.  (\cite{vlemmings02}) have measured the circular
polarization of the H$_2$O masers in three super-giants, NML\,Cyg,
VY\,CMa, and S\,Per, and the Mira variable U\,Her. The values obtained
for the magnetic field are 100--500\,mG for the super-giants and about
1\,G for the Mira variable. As we see the data are still scarce, but I
think that the potential of polarization measurements in probing the
magnetic field in AGB and post-AGB stars is clear, and that these type
of works will provide useful constrains for models of bipolar flow
launching at these stages of the stellar evolution.

\section{A look into the future}

Just from the few examples I just presented, I think that we can
conclude that the VLBI observations of circumstellar maser lines are
providing wonderful results with great impact not only on the maser
phenomenon itself and the physical and chemical processes of AGB
envelopes, but also on other basic fields in Astronomy (such as the
Astrometry of stars with its implications on the dynamics and stellar
population of the Galaxy).  However, with the use of the new
instrumentation that is now becoming available, or will become
available in the next coming years, it is easy to foresee an even more
brilliant future for VLBI circumstellar maser research.

The already operating networks of the HSA and GMVA will lower the
detection limit, especially in the 3\,mm band, allowing a much better
study of weak maser lines, such as $v$=3 $J$=1--0 $^{28}$SiO masers or
the HCN masers at 89\,GHz. The new 40\,m dish now being built in Yebes
will certainlly join these networks soon, helping in solving the
collecting area limitation of VLBI at mm wavelengths.  Also, the EVLA
plus the VLBA will be a perfect instrumental combination for observing
\mbox{$J$=1--0} SiO masers at 43 GHz, as it is the EVN plus MERLIN at
lower frequencies.

In principle, the developement of $e$EVN does not represent a
sensitivity improvement over the traditional way of VLBI
observing. However it will certainly offer an opportunity for easing
the planning and scheduling of the observing sessions. In particular,
having more than three EVN sessions per year will greatly improve the
reliability of proper motion measurements against the {\em Christmas
Tree} effect. Note that for the case of maser lines, relatively narrow
bands (as compared with continuum observations) are needed. Therefore,
setting up an $e$EVN network for line studies does not require so much
networking bandwidth, being much cheaper to operate.

The Japanese VERA project will represent a mayor breakthrough for
circumstellar maser studies. This will be a dedicated phase reference
array to measure annual parallax and proper motions of AGB stars,
observing their masers at 22\,GHz (H$_2$O) and 43\,GHz (SiO). In
addition to provide positions to an accuracy of 10\,$\mu$a.s., and
precise distances anywhere in the Galaxy, the project will also release
maps of these two maser lines in a large number of sources.

The Observatorio Astron\'omico Nacional (OAN) is engaged in the ALMA
project, that at its highest operative frequencies will provide spatial
resolutions up to 6 m.a.s. Its enormous collecting area will let us
accurately map many spectral lines in circumstellar envelopes, as well
as to locate the central star.  It is possible that combining ALMA and
VLBI observations of SiO masers at 3\,mm we could finally place the
star at the center of the maser rings.

Moving higher in frequencies, it is planned that the Herschel satellite
will fly sometime in 2007. One of the major tasks for the heterodyne
instrument on board, HIFI (in which the OAN is deeply involved too),
will certainly be the exploration of water lines everywhere.  In this
context, circumstellar envelopes offer a unique opportunity because of
their simple geometry and kinematics, and the wide range of physical
conditions they exhibit. This instrument however, will be unable to
resolve the emission from water in CEs, the analysis of the results
being largely dependent on their modeling. Here is where a good
knowledge of the different CE regimes, via interferometric/VLBI
observations of all OH, H$_2$O and SiO maser lines, may play a key
role.

Going even higher in frequencies, the two interferometric instruments
at the VLTI, VLTI-VINCI (near-IR) and VLTI-MIDI (mid-IR), will offer a
unique opportunity to study the inner parts of CEs.  VLTI-VINCI is able
to measure the size of the central star and monitor its changes in size
and temperature along the pulsation cycle. VLTI-MIDI will do a similar
work but for the dust formation layer. It is clear that the combination
of these two instruments with VLBI observations of SiO and 22\,GHz
water masers, will give us a superb picture of the central star and its
influence through the SiO-maser and water-maser/dust-formation layers
(see e.g. Wittkowski \& Boboltz \cite{vinci}).

And of course, we should nor forget the SKA project, that will
certainly change the way we think about VLBI.

\begin{acknowledgements}

This contribution would not have been possible without the help of the
VLBI group at the OAN. I appreciate very much J.-F. Desmurs and R.
Soria-Ruiz for their critical reading of the manuscript, and the
editors of this proceedings book for their patience. I also want to
thank my wife for her comprehension, and my kids for making such a
joyful noise around, during the writing of this contribution.  This
work has been financially supported by the Spanish DGI (MCYT) under
projects AYA2000-0927 and AYA2003-7584, and by the European
Commission's I3 Programme ``RADIONET", under contract No.\ 505818.

\end{acknowledgements}

\end{document}

%% file: page.tex
\setcounter{page}{169}